\documentclass[prd,twocolumn,nofootinbib,notitlepage,aps,tightenlines,preprintnumbers,amsmath,amssymb,amsfonts,showpacs,superscriptaddress]{revtex4-1}

\RequirePackage[colorlinks=true
,urlcolor=blue
,anchorcolor=blue
,citecolor=blue
,filecolor=blue
,linkcolor=blue
,menucolor=blue
,linktocpage=true
,pdfproducer=medialab
,pdfa=true
]{hyperref}

\usepackage{amsmath,amssymb,amsthm,amsfonts}
\usepackage{graphicx,tabularx}
\usepackage{float}
\usepackage{multirow}  
\usepackage{hyperref}
\usepackage{cancel}
\usepackage{comment}
\usepackage{physics}
\usepackage{lineno}

\usepackage{cleveref}
\crefname{section}{Sec.}{Secs.}
\crefname{figure}{Fig.}{Figs.}
\crefname{equation}{Eq.}{Eqs.}
\crefname{appendix}{Appendix}{Appendices}

\newcommand{\be}{\begin{equation} \begin{aligned}}
\newcommand{\ee}{\end{aligned} \end{equation}}
\newcommand{\HE}{$^4{\rm He}$}

\usepackage{xcolor}
\usepackage{lineno}

\RequirePackage[normalem]{ulem}

\newcommand{\orcid}[1]{\begingroup
  \hypersetup{hidelinks}\href{https://orcid.org/#1}{\includegraphics[width=10pt]{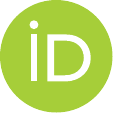}} \endgroup}

\begin{document}

\title{The X17 with Chiral Couplings}

\author{Max H. Fieg  \orcid{0000-0002-7027-6921}}
\email{mfieg@fnal.gov}
\affiliation{Department of Physics and Astronomy, University of California, Irvine, CA 92697 USA\\[0.1cm]}
\affiliation{Fermi National Accelerator Laboratory, Batavia, IL 60510 USA}

\author{Toni M{\"a}kel{\"a} \orcid{0000-0002-1723-4028}}
\email{tmakela@uci.edu}\affiliation{Department of Physics and Astronomy, University of California, Irvine, CA 92697 USA\\[0.1cm]}

\author{Tim M.P. Tait \orcid{0000-0003-3002-6909}}
\email{ttait@uci.edu}\affiliation{Department of Physics and Astronomy, University of California, Irvine, CA 92697 USA\\[0.1cm]}

\author{Mi{\v s}a Toman \orcid{0009-0003-2647-1281}}
\email{mtoman@uci.edu}\affiliation{Department of Physics and Astronomy, University of California, Irvine, CA 92697 USA\\[0.1cm]}

\begin{abstract}
In recent years, the ATOMKI collaboration has performed a series of measurements of excited nuclei, observing a resonant excess of electron-positron pairs at large opening angles compared to the Standard Model prediction. The excess has been hypothesized to be due to the production of a new spin-1 or spin-0 particle, X17, with a mass of about 17 MeV.  Recently, the PADME experiment has reported an excess in the $e^+e^-$ cross section at center-of-mass energies near 17 MeV, perhaps further hinting at the existence of a new state.
Studies of the spin-1 case have hitherto focused on either vector {\em or} axial-vector couplings to quarks and leptons, whereas UV theories more naturally produce {\em both} vector and axial-vector (\textit{i.e.} chiral) couplings, analogous to the Standard Model weak interactions. We consider the ATOMKI anomalies in the context of an $X$ with chiral couplings to quarks and explore the parameter space that can explain the ATOMKI anomalies, contrasting them with experimental constraints.
We find that it is possible to accommodate the reported ATOMKI signals.  However, the $99\%$ CL region is in tension with null results from searches for atomic parity violation and direct searches for new low mass physics coupled to electrons. This tension is found to be driven by the magnitude of the reported excess in the transition of $^{12}{\rm C}(17.23)$, which drives the best-fit region towards excluded couplings.

\end{abstract}\maketitle

\section{Introduction}

The Standard Model (SM) is a tremendously successful description of the electromagnetic, weak and strong nuclear forces. Nonetheless, it offers no explanation for a host of open questions, such as the origin of neutrino masses or the identity of dark matter. A common ingredient in theories seeking to address the open questions of the SM is the introduction of new particles, potentially mediating a previously unknown fifth force. 

While the SM has remained triumphant in most experimental tests in particle and nuclear experiments, occasionally a tension arises that stimulates discussion on the possibility of a new signal. One such instance is the famous ATOMKI anomaly~\cite{Krasznahorkay:2015iga}, an excess of electron-positron pairs that emerge from the decay of an excited beryllium nucleus with opening angles $\theta_{e^+ e^-} \approx 140\deg$. The result suggests a new boson $X$ hypothetically produced in the decay $^8$Be(18.15)$\to^8$Be$ + X$ and decaying predominantly to $e^- e^+$ pairs, at a mass of roughly $m_X \simeq 17$~MeV, hence termed the X17 particle. The ATOMKI collaboration has also rebuilt, re-performed and reanalyzed their experiment, confirming the anomaly~\cite{Krasznahorkay:2017bwh}, and made further measurements of $^8$Be~\cite{Krasznahorkay:2018snd, Krasznahorkay:2275461,Krasznahorkay:2017bwh}, $^{12}$C~\cite{Krasznahorkay:2022pxs}, and $^4$He~\cite{Krasznahorkay:2019lyl}. Similar results were obtained for $^8$Be$^*$ by the VINATOM experiment~\cite{Anh:2024req}, whereas the MEG II Collaboration observed no significant signal in an attempted reproduction of the original beryllium result~\cite{MEGII:2024urz}. 

Recently, the topic has been reinvigorated~\cite{Arias-Aragon:2025wdt,Cederkall:2025bka}
by the observation of the PADME positron annihilation experiment, entirely dissimilar to the nuclear experiments, yet hinting at an excess compatible with the X17~\cite{Bossi:2025ptv}.
Candidates for the X17 have been discussed e.g. in Refs.~\cite{Feng:2016jff, Kozaczuk:2016nma, Feng:2016ysn,Feng:2020mbt,Barducci:2022lqd,Kozaczuk:2016nma,Mommers:2024qzy}, and pseudoscalar models have been deemed incompatible with the $^{12}$C measurements~\cite{Krasznahorkay:2022pxs}. Nevertheless, much of the previous work has focused on bosons with purely vector or axial couplings, hence not encompassing the full set of possibilities and in contrast to the Standard Model. 

The present work considers a vector boson $X_\mu$ with chiral couplings to isospin violating and conserving nuclear axial-vector and vector currents. The relevance of the present study is further increased by the constraints on vector and axial-vector $X$ set by null results from other searches \cite{Kahn:2016vjr}.  Ref.~\cite{Hostert:2023tkg} found that the measurements of rare pion decay at the SINDRUM-I spectrometer rule out an X17 explanation with purely vector couplings and strongly constrain the axial-vector couplings. These constraints can be evaded if the $X$ has an electron coupling that is related to the quark couplings so as to cancel the amplitude to produce $X$ in pion decay. The consequence of this is a tuned electron coupling, subject to constraints from searches for new physics coupled to electrons, the most relevant of which are KLOE-2~\cite{Anastasi:2015qla} and NA64~\cite{NA64:2018lsq,NA64:2019auh}.

While the $X$ with chiral couplings represents a compelling candidate to explain the ATOMKI anomalies, we find that our model cannot simultaneously fit all measurements and evade constraints at face value. Our best-fit region to the ATOMKI measurements at 99\% CL is ruled out by a combination of searches for new physics at KLOE-2, and results in a sizable parity violation that should have been observed in Cesium systems. This tension is largely driven by the magnitude of the ATOMKI $^{12}{\rm C}(17.23)$ signal, for which theoretical predictions require knowledge of the currently poorly-determined axial-vector matrix element.  We find that transitions involving $^{12}{\rm C}(17.23)$ are a topic where future experimental observation and/or improved theoretical modeling could do much to clarify the current situation.

The work is structured as follows. 
\cref{sec:methodology} discusses the methodology of the present work, while \cref{sec:existing_constraints} introduces the experimental constraints. The results are presented in \cref{sec:results}, and conclusions are drawn in \cref{sec:conclusions}. The experimental measurements from the ATOMKI collaboration are reviewed in \cref{app:ATOMKI_review}, while \cref{app:EFTCalc} details the effective field theory formalism for calculating the partial width of the decay of each nuclear state to $X$. 

\section{Formalism and Methodology}
\label{sec:methodology}

The formalism for relating the underlying coupling to quarks to the relevant nuclear transitions is presented in detail in \cref{app:EFTCalc} and in Ref.~\cite{Feng:2020mbt}.  Here, we briefly summarize and discuss important prior constraints. We extend the SM by a vector field $X_\mu$ with chiral couplings to isospin-violating and -conserving nucleon (axial-)vector currents of quarks (see \cref{eq:currentsv0,eq:currentsv1,eq:currentsa0,eq:currentsa1}). At the nucleon-level, the interactions read
\begin{equation}
\label{eq:lagrangian}
\begin{aligned}
    {\cal L}_X = \tfrac{1}{2} X^\mu \big[ & (e\epsilon_p^V + e\epsilon_n^V) J_\mu^0 
    + (e\epsilon_p^V - e\epsilon_n^V) J_\mu^1 \\
    &+ (\epsilon_p^A + \epsilon_n^A) J_{\mu,5}^0 
    + (\epsilon_p^A - \epsilon_n^A) J_{\mu,5}^1 \big],
\end{aligned}
\end{equation}
where $\epsilon_N^{V(A)}$ is the (axial-)vector coupling to nucleon $N\in\{p,n\}$, normalized to the electric charge $e$ for the vector coupling. $J^{0,1}_{\mu(5)}$ are the isospin- and parity- conserving (changing) currents. The interactions in \cref{eq:lagrangian} map on to the amplitudes for the nuclear transitions via $X$ emission. These further depend on nuclear matrix elements of the form $\langle N|J_\mu |N_*\rangle$, with $N_{(*)}$ denoting the (excited) nuclear state.

Following Ref.~\cite{Feng:2020mbt}, the decay of each of the nuclei is described in terms of a nuclear effective field theory (EFT) built from a set of operators subject to parity and angular momentum conservation, with separate operators encapsulating the coupling through axial-vector and vector currents. 
For example, the operators mediating the exotic $1^+ \rightarrow 0^+$ beryllium decay are
\begin{align}
\label{eq:BeOperator_text}
{\cal O}_5 &= \frac{C_V^{\rm Be}}{\Lambda} N_0^\dagger 
\epsilon^{\mu\nu\alpha\beta}\partial_\mu N^*_{\nu}\partial_\alpha X_\beta, \nonumber \\
{\cal O}_3 &= C_A^{\rm Be} \Lambda N_0^\dagger N^*_\mu X^\mu,
\end{align}
where the first (second) term is the (axial-)vector operator, and the $C^{\rm Be}_{V,A}$ are the Wilson coefficients (with common dimensionful scale $\Lambda$ extracted for clarity), related to the matrix elements by matching the EFT to the nucleon level amplitudes. These matrix elements can be estimated using  various approaches. 

The Wilson coefficients for vector operators are calculated by relating the partial decay width to $X$ to that of the partial decay width of the photon, making use of the fact that both processes involve combinations of the same currents $J^{0,1}_{\mu}$. Additional work is required for helium, for which the relevant excited state does not decay to photons, but to electron-positron pairs through an $E0$ transition -- these can be related to the operator mediating the decay to $X$ through the equations of motion~\cite{Feng:2020mbt}. The same tactic does not work for axial-vector operators, and determination of the necessary matrix elements  relies on estimates from nuclear theory. The beryllium axial matrix element was calculated using $N$-body techniques of nuclear structure~\cite{Kozaczuk:2016nma}; the helium matrix element was reported in Ref.~\cite{Horiuchi:2009zza} and applied to the X17 in Ref.~\cite{Barducci:2022lqd}; and the carbon axial matrix element was recently estimated using a nuclear shell model~\cite{Mommers:2024qzy}.

No uncertainty is quoted for the helium matrix element, and the carbon calculation shows some tension with existing measurements, as discussed at the end of this paragraph. Consequently, the uncertainties on both helium and carbon axial matrix elements are under less control than those for beryllium. We deal with this situation by assigning benchmark uncertainties of $50\%$ and $100\%$, and our results are presented separately for the two cases. However, we note that this treatment of the uncertainty in the axial-vector matrix elements may not be sufficient, particularly for the case of carbon. In Ref.~\cite{Mommers:2024qzy}, the decay width of $^{12}{\rm C}(17.23)$ to an axial-vector $X$ was calculated and related to the calculation for the SM decay to a photon --- but the prediction for the latter process was found to be a factor of $\approx 5.4$ larger than the experimentally measured value. Given this deviation, we also consider the case where the axial-vector contribution to the partial decay width for the exotic decay is a factor of 5 larger or smaller than the central value.  Ultimately we find that even this expanded uncertainty does does not significantly change our conclusions.

\section{Measurements}
\label{sec:existing_constraints}

\begin{table}[h!]
\centering
\begin{tabular}{l | c}
\hline
\textbf{Nucleus} & \textbf{ATOMKI Measurement} \\ 
\hline
$^8{\rm Be}(18.15)$ & ${\rm B}_X=(6\pm1)\times 10^{-6}$~\cite{Krasznahorkay:2018snd} \\ 
$^8{\rm Be}(17.64)$ & ${\rm B}_X=(2\pm 2)\times 10^{-6}$~\cite{Krasznahorkay:2017bwh} \\ 
$^{12}{\rm C}(17.23)$ & ${\rm B_X}=(3.6\pm 0.3)\times 10^{-6}$~\cite{Krasznahorkay:2022pxs}  \\ 
$^4{\rm He}(20.21),^4{\rm He}(21.01)$ & $\sigma_X/\sigma_{\rm E0} = 0.2$~\cite{Krasznahorkay:2019lyl} \\ 
\hline

\end{tabular}
\caption{ATOMKI measurements and experimental uncertainties.  ${\rm B}_X=\Gamma_X/\Gamma_\gamma$ is the ratio of the partial decay width to $X$ to that of the photon, assuming $100\%$ $X \to e^- e^+$ branching fraction. For helium, the decay is mediated from a mixture of excited states $^4{\rm He}(20.21),^4{\rm He}(21.01)$, and so the relevant observable is the total cross section for $X$, $\sigma_X$ normalized to the SM production for the ${\rm E0}$ transition, $\sigma_{\rm E0}$. No uncertainty is presented for the $\sigma_X/\sigma_{\rm E0}$ measurement; the same relative experimental uncertainty as ${\rm B}_X(^8{\rm Be}(18.15))$ is assumed.\label{tab:ATOMKImeasurements}}
\end{table}

The measurements by the ATOMKI collaboration considered here are summarized in \cref{tab:ATOMKImeasurements}, and detailed in \cref{app:ATOMKI_review}. For each of the listed excited nuclei, the ATOMKI collaboration reported the best-fit ${\rm B}_X$, which is the branching ratio of the excited state to $X$, divided by its decay to a (off-shell, in the case of helium) photon. For the helium measurement, a proton beam energy was used to produce an excitation intermediate between the parity even and parity odd states, $^4{\rm He}(20.21),^4{\rm He}(21.01)$. Thus for helium the relevant experimental observable is the $X$ production cross section, expressed as a ratio to the ${\rm E0}$ transition cross section, $\sigma_x/\sigma_{\rm E0}$. No experimental uncertainty was reported for the $^4$He 
measurement.  We assume a relative experimental uncertainty equal to the relative error quoted for the $^8{\rm Be}(18.15)$ measurement. It is worth mentioning that excesses possibly originating from $X$ were also observed in the giant dipole resonance transition of beryllium by ATOMKI~\cite{Krasznahorky:2024adr,Krasznahorkay:2023sax}, and in the $^8{\rm Be}(18.15)$ decay by the VNU collaboration~\cite{Anh:2024req}. However, unless otherwise stated, we do not include these measurements in our fits.

Each ATOMKI measurement, and the recent PADME result~\cite{Bossi:2025ptv}, have a best-fit for the mass pointing to $m_X\approx 16.8~{\rm MeV}$. Ref.~\cite{Arias-Aragon:2025wdt} performed a combined fit for the $X$ mass, including the ATOMKI, MEG-II, VNU, and PADME measurements, producing a $\Delta\chi^2$ profile as a function of $m_X$. All measurements point to an $X$ with a mass of 17 MeV within $1\sigma$.

Besides the ATOMKI signals, there are a number of important constraints from null results for searches for the $X$. The MEG-II~\cite{MEGII:2018kmf} experiment searched for $X$~\cite{MEGII:2024urz} by producing the excited beryllium state and searching for the $X\to e^+e^-$ decay. No excess was observed, and limits were set on ${\rm B}_X$ for the $^8{\rm Be}(18.15)$ and $^8{\rm Be}(17.64)$ states. These have some tension (about $1.5\sigma$ for the $^8{\rm Be}(18.15)$ transition) with the best-fit from ATOMKI. The tension in the $^8{\rm Be}(17.64)$ transition is stronger. 

Rare pion decays at SINDRUM-I~\cite{Eichler:2021efn} have been shown to provide stringent constraints on light vector bosons~\cite{Hostert:2023tkg}. At 90\% CL, these bounds can be expressed as\footnote{The quark vector couplings are related to the nucleon couplings by $\epsilon^V_{p(n)}=2(1)\epsilon_u^V+(2)\epsilon_d^V$, and the axial-vector couplings to the spin contribution $\epsilon_{p(n)}^A=\Delta u^{p(n)}\epsilon_u^A + \Delta d^{p(n)}\epsilon_d^A$, where $\Delta u^{p} = \Delta d^{n} = 0.897 $ and $\Delta u^{n} = \Delta d^{p}=-0.367$\cite{Kozaczuk:2016nma, Hostert:2023tkg}.} 
\begin{equation}
\begin{split}
    e\epsilon|\Delta Q_X| &= |(e\epsilon_u^V+\epsilon_u^A) - (e\epsilon_d^V+\epsilon_d^A) \\
    &\quad - (e\epsilon_\nu^V-\epsilon_\nu^A) - (e\epsilon_e^V-\epsilon_e^A)|
    < 8.5\times10^{-5}
\end{split}
\end{equation}
for 100\% branching fraction of $X$ to electron-positron pairs~\cite{Hostert:2023tkg}. 
The coupling to electrons is largely unconstrained by ATOMKI with the only requirement being that it results in a prompt decay (which is satisfied for $\epsilon_e^{V,A}\gtrsim 10^{-5}$). When fitting the calculated widths to the ATOMKI measurements, we fix the electron vector coupling to cancel the SINDRUM-I constraint, via
\begin{equation}
\label{eq:electronvector}
    e\epsilon_e^V= (e\epsilon_u^V+\epsilon_u^A) - (e\epsilon_d^V+\epsilon_d^A), \quad
\end{equation}
with the coupling of $X$ to neutrinos and its axial coupling to the electron taken to be zero.

Light spin-1 bosons coupled to electrons face further constraints from various low energy beam dumps, such as KLOE-2, constraining $\sqrt{(e\epsilon_e^V)^2 +  (\epsilon_e^A)^2  } < 2\times 10^{-3}e$, and the NA64 experiment which requires $\epsilon_e^V>6.8\times 10^{-4}$\cite{NA64:2018lsq,NA64:2019auh}.

Similarly, vector couplings to nucleons face a multitude of constraints. Rare pion decays at NA48~\cite{NA482:2015wmo} constrain the vector couplings to quarks, resulting in the so-called ``protophobic" bound $|\epsilon_p^V| = |2\epsilon_u^V+\epsilon_d^V|<8\times 10^{-4}$~\cite{Feng:2016jff}. Searches for long-range interactions between neutrons and nuclei bound the neutron vector coupling to $|\epsilon_n^V|<2.5\times 10^{-2}$~\cite{Feng:2016jff}. 

Axial couplings to nucleons are relatively less constrained. However, because we consider models with mixed axial and vector couplings, there are powerful constraints from searches for parity violation~\cite{Porsev:2009pr}. In Ref.~\cite{Dzuba:2017puc}, generic constraints were derived for light vector bosons with chiral couplings, and it was found that for vector bosons with mass $\lesssim 100 ~{\rm MeV}$, measurements from Cesium atoms constrain,  
\begin{equation}
\label{eq:PV}
    |\epsilon_p^A\epsilon_{\rm Cs}^V|~<~6\times10^{-8}
\end{equation}
where $\epsilon_{\rm Cs}^V =e(Z_{\rm Cs}\epsilon_p^V + N_{\rm Cs}\epsilon_n^V )/A_{\rm Cs}$.

\section{Results}
\label{sec:results}

\begin{figure*}[ht!]
    \centering
    \includegraphics[width=0.95\linewidth]{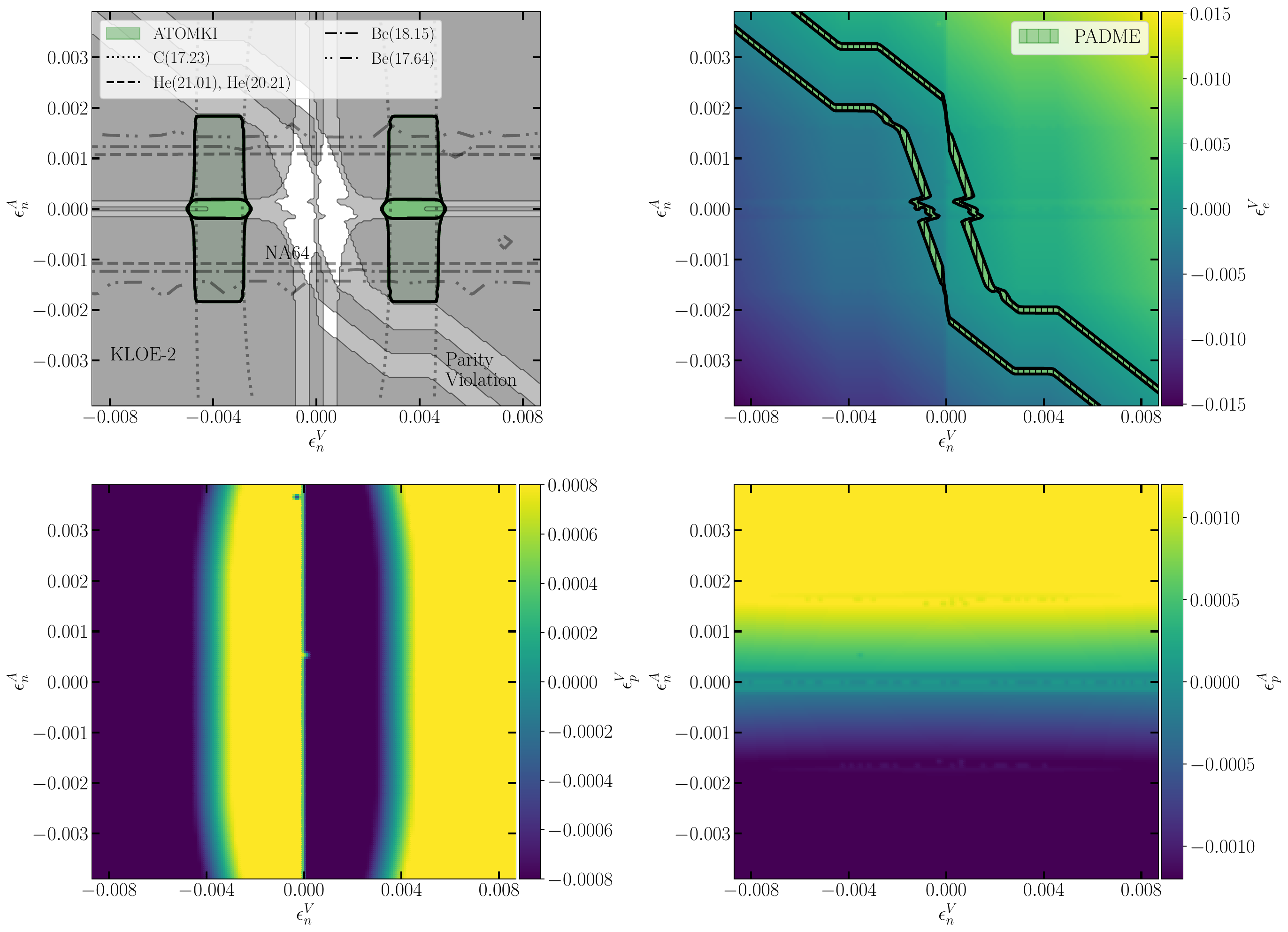}
    \caption{The result of the fit to the ATOMKI measurements in \cref{tab:ATOMKImeasurements}. For each point in the plane of the neutron couplings $\epsilon_n^V, \epsilon_n^A$, we determine the proton couplings $\epsilon_p^V, \epsilon_p^A$ best fitting the ATOMKI measurements, and the electron coupling $\epsilon_e^V$ canceling the constraints from SINDRUM-I. In the upper left plot, the bright green region with a solid black outline indicates agreement with ATOMKI at 99\% CL; the opaque dark green region with a dotted outline uses the larger theory uncertainty on the carbon and helium axial matrix elements as discussed in the text.  Regions compatible with individual ATOMKI measurements are shown by single dot-dashed lines ($^8{\rm Be}(18.15)$), double dot-dashed lines ($^8{\rm Be}(17.64)$), dashed lines ($^4{\rm He}$) and dotted lines ($^{12}{\rm C}(17.23)$).  The gray region denotes exclusion regions driven by parity violation constraints and the NA64, KLOE-2 experiments. The other three panels show the corresponding values of $\epsilon_e^V$ (upper right), $\epsilon_p^V$ (lower left), and $\epsilon_p^A$ (lower right) at each point in the plane of the neutron couplings.  The two green bands in the upper right panel indicate the $|\epsilon_e^V|$ necessary to to explain the PADME excess.
    }
    \label{fig:ProjectionPlot}
\end{figure*}

Having set the neutrino and electron axial couplings to the $X$ to zero, the parameter space consists of the six quantities, including the proton couplings $e\epsilon_p^V,\epsilon_p^A$, neutron couplings $e\epsilon_n^V,\epsilon_n^A$, electron coupling $e\epsilon_e^V$ and the mass $m_X$.  We scan the plane of the neutron couplings $\epsilon_n^V,\epsilon_n^A$, and for each fixed point, we fit the proton couplings (within the protophobic bound from NA48)
and $m_X$ to best fit the measurements in \cref{tab:ATOMKImeasurements} and the (combined) likelihood profile for $m_X$ \footnote{The analysis from Ref.~\cite{Arias-Aragon:2025wdt} does not report a global best-fit, so we take $\Delta \chi^2=\chi^2$. Note that this neglects some small tension in the different measurements of $m_X$.} from Ref.~\cite{Arias-Aragon:2025wdt}.
Due to the high sensitivity of the likelihood profile for $m_X$, the mass essentially does not deviate from 16.9 MeV in the fits.  For each point in the scan, we adjust the electron vector coupling according to \cref{eq:electronvector} in order to cancel the SINDRUM-I constraint.  The best fit values of $\epsilon^V_e$,
$\epsilon^V_p$, and $\epsilon^A_p$ corresponding to each point in the plane of the neutron couplings are shown in the upper right, lower left, and lower right panels of \cref{fig:ProjectionPlot}, respectively. While we do not present the specific results here, we find that similar scans over different combinations of couplings lead to very similar conclusions.

The regions of the scan best matching the ATOMKI observations are shown in the upper left panel of \cref{fig:ProjectionPlot}.  These regions strongly depend on the assumed theoretical uncertainties on the axial vector nuclear matrix elements for $^{12}$C and $^4$He.  For the most conservative choice of  $\pm 50\%$ uncertainty on those matrix elements, there is no overlap of regions at the $95\%$ CL, but regions consistent with all three measurements begin to appear at the $99\%$ CL, indicated by the two regions enclosed by the solid black lines and with bright solid green interior shading.  These regions favor vector $X$ coupling to neutrons around $\pm  0.004$, smaller axial vector couplings to neutrons, and protophobic vector and axial-vector couplings.  Also shown for reference are the regions favored at $1\sigma$ for the observation of each nuclear transition individually; evident from the plot is the fact that internal tension within the set of ATOMKI measurements drives the relatively small region compatible with all of them, as much wider regions of parameter space can provide better fits to each individual measurement.  The comparison of the best-fit region and those from the individual measurements also reveals the special role that the $^{12}$C observation plays in driving the parameter space for the couplings by demanding a relatively large $|\epsilon_n^V|$.  If one assumes the uncertainty on the carbon and helium axial vector matrix elements is increased to $\pm 100\%$, the best fit region expands to the opaquely shaded dark green region, extending the best-fit region in the direction of the neutron axial-vector coupling.

As mentioned above, we have chosen to tune the electron vector coupling to exactly cancel the strong constraints from the SINDRUM-I experiment. The upper right panel of ~\cref{fig:ProjectionPlot} shows the resulting value of $\epsilon_e^V$, which is generally around $|\epsilon_e^V|\sim{\cal O}(10^{-2})$, roughly the same size required to produce the excess observed by the PADME experiment. The green solid bands with hatches show the region where the observed limit on $\epsilon_e^V$ from PADME exceeds the expected limit by $1\sigma$ for $m_X\approx 17$ MeV, and $4.6\times 10^{-4}\lesssim |e\epsilon_e^V|\lesssim 5.6\times 10^{-4}$. The region between the two bands shows where the electron coupling is too small to match the excess, while in the region outside, it is too large.
Given the results of the ATOMKI fit and corresponding electron coupling tuned to cancel SINDRUM-I, constraints on $\epsilon_e^V$ can be projected onto the plane of the neutron couplings.  The regions excluded by the KLOE-2 and NA64 experiments and constraints from parity violation in Cesium atoms are shown in the upper left panel by gray shading.  They rule out\footnote{It is worth mentioning that our results do not exclude the possibility of regions marginally consistent with both ATOMKI and experimental constraints, because we do not perform a {\em combined} likelihood fit including both ATOMKI and the constraints.  That said, our results clearly establish  significant tension between the two.} the best-fit region at the $95\%$ CL. The white region near the middle of the plot denotes the combination of couplings that evade the constraints.  These constraints track the electron coupling, which the experiments bound from above and below, respectively. The parity violation constraints are the result of \cref{eq:PV}, with allowed regions corresponding to either $\epsilon_p^A$ or $\epsilon_{\rm Cs}^V$ becoming sufficiently small. 

\begin{figure*}[ht!]
    \centering
    \includegraphics[width=0.95\linewidth]{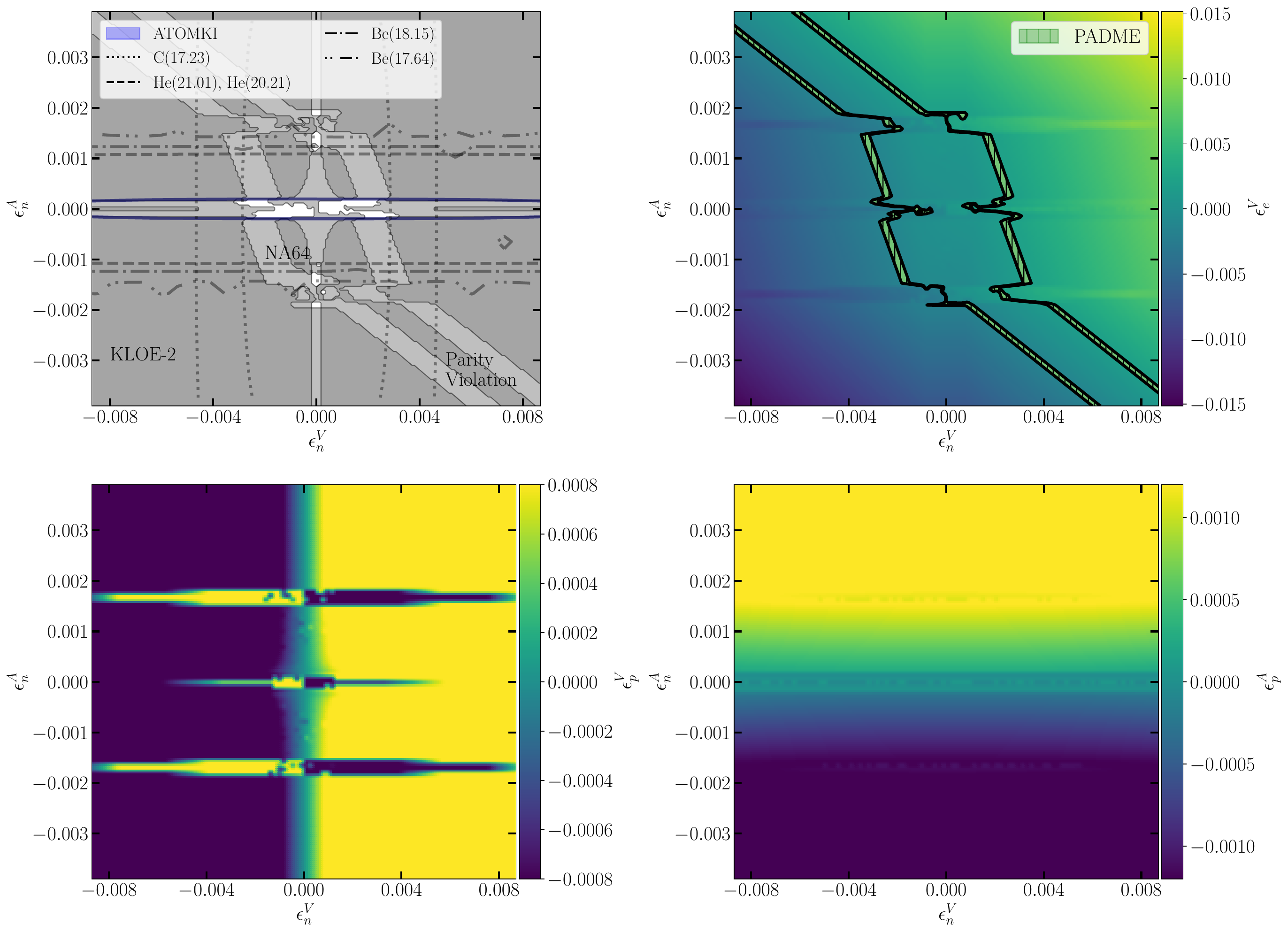}
    \caption{The result of the fit to the ATOMKI measurements in \cref{tab:ATOMKImeasurements} with $^{12}$C(17.23) excluded from the fit. In the upper left panel, the blue shaded region shows where the ATOMKI beryllium and helium observations are consistent at the $99\%$ CL.  In the upper left panel, the region outlined in blue shows where the ATOMKI beryllium and helium observations are consistent at the $99\%$ CL.
    }
    \label{fig:ProjectionPlot2}
\end{figure*}

We examine the importance of the ATOMKI $^{12}$C(17.23) measurement in driving these conclusions by reproducing the analysis of \cref{fig:ProjectionPlot}, but excluding the ATOMKI $^{12}$C(17.23) observation from the fit to the proton couplings.  The results are shown in \cref{fig:ProjectionPlot2}, and indicate that without the carbon measurement, the region of $99\%$ CL consistency expands to the blue band shown in the upper left panel, and includes regions unconstrained by KLOE-2, NA64, and searches for atomic parity-violation.  The best fit region favors vector couplings to the proton and the neutron, with the proton coupling close to saturating the protophobic bound.  We conclude that the size of the $^{12}$C transition is indeed a substantial obstacle to realizing parameters that are not excluded, but which fit all of the ATOMKI observations. 

\section{Conclusions}
\label{sec:conclusions}

The origins of the ATOMKI observations are yet to be understood, with the most exciting possibility being that a new particle $X$, with a mass of about 17 MeV, is produced in the decay of the excited nucleus, subsequently decaying into electrons. One of the leading hypotheses for the nature of $X$ is a spin-1 particle with either vector or axial-vector couplings to quarks and electrons.  While couplings conserving parity simplify consideration of experimental constraints, it remains possible that $X$ has chiral couplings to fermions, as in the case of the electroweak spin-1 particles of the SM. 

We consider an $X$ that has chiral couplings to quarks and electrons. We make predictions and fit to the ATOMKI measurements to assess whether this explanation can satisfy all of the ATOMKI observations as well as existing constraints from other experiments. For example, a spin-1 particle with chiral couplings is subject to constraints arising from searches for parity violation, which do not apply for a vector with purely vector or axial-vector couplings. A particularly powerful constraint comes from rare pion decays, depending on a linear combination of quark and electron couplings, the latter of which are not well-constrained by ATOMKI. We find that these constraints can thus be satisfied by adjusting the electron coupling to cancel out the quark contribution, and explore this possibility.  We choose to set the electron axial-vector couplings to zero, as otherwise the parity violation constraints become much stronger.

We find that there is a region of parameter space where ATOMKI measurements from beryllium, helium, and carbon transitions can be satisfied at the $99\%$ CL. However, the best-fit parameter space where all predictions overlap is ruled out by constraints from parity violation, and is additionally in tension with KLOE-2.
This tension is largely driven by the ATOMKI measurement of carbon, which demands a large vector coupling to neutrons. If this measurement is removed from the fit, an open region of parameter space is found for the remaining ATOMKI measurements considered. Moreover, the recent PADME measurement reports an excess in the electron-positron cross section for $\sqrt{s}\approx 17~{\rm MeV}$ and, under the $X$ hypothesis, points to an electron coupling of $|e\epsilon_e|\approx5.6\times10^{-4}$. Assuming this to be purely a vector coupling, we find that there is some open parameter space where the ${}^8{\rm Be}(18.15),^8{\rm Be}(17.64)$ and ${}^4{\rm He}$ are all satisfied and agree with all considered constraints.

That said, we are aware of no compelling physical reason to exclude $^{12}$C from the fit.  One could imagine that the failure of the shell model to accurately describe the electromagnetic transition is perhaps symptomatic that the $^{12}$C(17.23) state is not well-described as a $1^-$ resonance\footnote{Such a development would also re-open the possibility that the X17 is a pseudo-scalar.}.  Regardless, new experimental observations of the transition and/or improved nuclear modeling could dramatically clarify the situation.

Further investigations for the origin of the ATOMKI excesses are clearly needed. The $X$ must couple to electrons, and future PADME measurements with greater integrated luminosity can either confirm or refute the new particle interpretation. More detailed studies on the matrix elements for helium and carbon would help place the  mapping from nucleon-level couplings to the corresponding nuclear transition rates on firmer footing. Should future measurements firmly establish a self-consistent picture for the X17 that includes chiral couplings, a dedicated model-building effort would be warranted to address the challenges of realizing self-consistent UV theory consistent with all experimental constraints.

\section*{Acknowledgments}

We thank Jonathan Feng for useful discussions.
The work of M.F.~was supported in part by U.S. National Science Foundation Grants PHY-2111427 and PHY-2210283, as well as the Fermi Forward Discovery Group, LLC under Contract No. 89243024CSC000002 with
the U.S. Department of Energy, Office of Science, Office of High Energy Physics.
The work of T.M. is supported in part by U.S.~National Science Foundation Grants PHY-2111427 and PHY-2210283 and Heising-Simons Foundation Grant 2020-1840.
The research activities of TMPT are supported in part by the U.S. National Science Foundation under Grants PHY-2210283 and PHY-2514888.
The work of M.T. is supported in part by U.S. National Science Foundation Grants PHY-2111427 and PHY-2210283 and PHY-2514888.

\appendix

\section{ATOMKI Measurements}
\label{app:ATOMKI_review}

All measurements are roughly consistent with a spin-1 particle with axial-vector, vector, or mixed couplings to quarks, with a mass of $m_X\approx 17 ~{\rm MeV}$ and decaying primarily to an electron-positron pair. For extended discussion, cf. Refs.~\cite{Feng:2020mbt,Barducci:2022lqd}.

\subsection{Beryllium} 

There are two relevant $1^+$ excited states with comparable energy levels: a mostly isoscalar state $^8{\rm Be}$(18.15) and a mostly isovector state $^8{\rm Be}$(17.64) (there is isospin mixing and breaking of the nearby states, as we will discuss in more detail). The anomaly was first measured and reported for the isoscalar state ~\cite{Krasznahorkay:2015iga}, with the best fit for the ratio of the partial width of the $X$ to that of the photon multiplied by the branching fraction of $X$ to an electron positron pair to be~\cite{Krasznahorkay:2018snd}
\begin{equation}
\label{eq:Beratioisoscalar}
    \frac{\Gamma(^8{\rm Be}(18.15)\to {}^8{\rm Be}+X)}{\Gamma(^8{\rm Be}(18.15)\to {}^8{\rm Be}+\gamma)} {\rm BF}(X\to e^+ e^-) = 
    (6\pm 1)\times10^{-6}.
\end{equation}
The isovector state was not observed to have any anomalous decays in the original search, but was later observed to exhibit a decay consistent with a $\approx 17 $ MeV boson in Refs.~\cite{Krasznahorkay:2275461,Krasznahorkay:2017bwh}. The measurement was reported as 
\begin{equation}
    \frac{\Gamma(^8{\rm Be}(17.64)\to {}^8{\rm Be}+X)}{\Gamma(^8{\rm Be}(17.64)\to{}^8{\rm Be}+\gamma)} {\rm BF}(X\to e^+ e^-) = (2 \pm 2)\times10^{-6}.
\end{equation}
Note that an exotic decay for the $^8{\rm Be}(17.64)$ state is expected to have additional phase space suppression as compared to the $^8{\rm Be}(18.15)$.

The exotic decay of the $^8{\rm Be}(18.15)$ was searched for at the electron-positron pair spectrometer at the VNU University of Science\cite{Anh:2024req} which found a statistically significant excess above background. They report a ratio of decay widths $\Gamma_X^{\rm Be(18.15)}/\Gamma_\gamma^{\rm Be(18.15)}=1.1\times10^{-5}$, which is about twice the value found by ATOMKI \cref{eq:Beratioisoscalar}.

\subsection{Helium} 

\HE ~has two relevant excited states. A parity even \HE(20.21) state with $J^P=0^+$ and a parity odd \HE(21.01) with $J^P=0^-$. An $X$ with mixed vector and axial-vector couplings to quarks could be produced in the decay of either excited state. The ATOMKI collaboration ran the experiment at an intermediate energy, so that both states would comparably contribute. The cross section for $X$ production, $\sigma_X$, is calculated by relating the production cross section for the parity odd (even) state $\sigma_-$ $(\sigma_+)$ to the total decay width $\Gamma_-$ ($\Gamma_+$) for each state, and normalizing to the E0 internal pair conversion cross section ~\cite{Feng:2020mbt}
\begin{equation}
    \frac{\sigma_X}{\sigma_{\rm E0}} = \frac{\Gamma(0^+\to X)}{\Gamma_{\rm E0}} + \frac{\sigma_-\Gamma_+}{\sigma_+\Gamma_-}\frac{\Gamma(0^-\to X)}{\Gamma_{\rm E0}},
\end{equation}
where $\Gamma(0^\pm\to X)$ is the partial decay width to each state. The ATOMKI collaboration has reported~\cite{Krasznahorkay:2019lyl} a measurement of $\sigma_X/\sigma_{\rm E0}=0.2$, with no uncertainty provided. Following Ref.~\cite{Barducci:2022lqd}, we assign a relative experimental uncertainty on $\sigma_X/\sigma_{\rm E0}$ equal to that of the beryllium isoscalar. 

\subsection{Carbon} 

The ATOMKI collaboration also observed an excess in the decay of the $1^-$ excited state of carbon, $^{12}$C(17.23)~\cite{Krasznahorkay:2022pxs}. The best fit width of the excited state decaying to $X$, multiplied by the $X\to e^+ e^-$ branching fraction and normalized to the partial decay to photons, was reported to be
\begin{equation}
\begin{aligned}[t]
\frac{\Gamma(^{12}{\rm C}(17.23)\!\rightarrow\!^{12}{\rm C}+X)}
     {\Gamma(^{12}{\rm C}(17.23)\!\rightarrow\!^{12}{\rm C}+\gamma)}
\, {\rm BF}(X\!\rightarrow\! e^+ e^-) \\[1pt]
= (3.6 \pm 0.3) \times 10^{-6}.
\end{aligned}
\end{equation}

\section{Nuclear EFT Calculations}
\label{app:EFTCalc}

In this section we discuss the construction of the EFT describing the $X$ contribution to nuclear transitions, under the hypothesis that $X$ is a spin-1 particle with mixed vector / axial-vector couplings to quarks, following the formalism developed in Ref.~\cite{Feng:2020mbt}. The starting point is the $X$ interactions with nucleons via isospin conserving and violating currents with (axial-)vector couplings, $J_{\mu(,5)}^0,J_{\mu(,5)}^1$. The nucleon EFT is then matched to a nucleus level EFT from which nucleus matrix elements can be extracted. In particular, the $X$ couples to nucleons via vector couplings $X_\mu\left(1/2(\epsilon_p^V + \epsilon^V_n)e \ J^\mu_0 + 1/2(\epsilon_p^V - \epsilon^V_n)e \ J^\mu_1    \right)/2$, where the first (second) term mediates isospin (non)conserving transitions. The $\epsilon^V_p$ ($\epsilon^V_n$) is the associated vector coupling to protons (neutrons). The axial-vector coupling case is similar, with $e\epsilon^{V}_{p,n}\to \epsilon^A_{p,n}$ and $J_\mu^{0(1)}\to J_{\mu,5}^{0(1)}$. 

The nucleon currents are defined as
\begin{align}
\label{eq:currentsv0}
J_\mu^0 &= \bar{p}\gamma_\mu p + \bar{n}\gamma_\mu n, \\
\label{eq:currentsv1}
J_\mu^1 &= \bar{p}\gamma_\mu p - \bar{n}\gamma_\mu n, \\
\label{eq:currentsa0}
J_{\mu,5}^0 &=\bar{p}\gamma_\mu\gamma^5 p + \bar{n}\gamma_\mu\gamma^5 n, \\
\label{eq:currentsa1}
J_{\mu,5}^1 &= \bar{p}\gamma_\mu\gamma^5 p - \bar{n}\gamma_\mu\gamma^5 n.
\end{align}

At the nucleus level, operators are constructed based on the spin-parity of $X$, the excited and ground states of each nucleus. The relevant operators for the (axial-) vector couplings of $X$ to each of the nuclei are given in Table III of Ref.~\cite{Feng:2020mbt}. The benefit of this EFT is that Wilson coefficients can be matched to nuclear matrix elements obtained either by measurements of the nuclei decaying to a photon for vector couplings, or through nuclear modeling techniques for axial-vector couplings.

\subsection{Beryllium} 

The leading nucleus EFT operator contributing to the exotic decay of the mostly iso-vector state, ${\rm Be}(17.64)$, is 
\begin{equation}
\label{eq:BeOperator}
    {\cal O}_5=C_V^{\rm Be}\Lambda^{-1} N_0^\dagger \epsilon^{\mu\nu\alpha\beta}\partial_\mu N^*_{\nu}\partial_\alpha X_\beta, ~{\cal O}_3= C_A^{\rm Be}\Lambda N_0^\dagger N^*_\mu X^\mu,
\end{equation}
where the first term is the vector piece, the second is the axial-vector, $N_0 ~(N^*)$ is the ground (excited) state and $\Lambda$ is a scale with dimensions of energy. The amplitude for the decay of the excited state then reads ${\cal M}={\cal M}_V+{\cal M}_A=C_V^{\rm Be}\Lambda^{-1} \epsilon^{\mu\nu\alpha\beta}p^*_\mu \epsilon^*_{\nu} p_\alpha \epsilon^X_\beta + C_A^{\rm Be}\Lambda \epsilon^*_\mu \epsilon^{X^\mu}$. Averaging and summing over the initial and final spins of the vector states, the squared-amplitudes do not interfere, since $\overline{{\cal M}_V{\cal M}_A}\propto\left(\epsilon^{\mu\nu\alpha\beta}g_{\nu\rho}g_{\beta\rho}\right)p^*_\mu p^X_\alpha=0$. Hence the resulting total width of the particle is simply the sum of the partial widths for the vector and axial-vector couplings, \textit{i.e.} $\Gamma_X^{\rm Be} = \Gamma_{X,V}^{\rm Be} + \Gamma_{X,A}^{\rm Be}$.
    
The partial widths are written as
\begin{equation}
\label{eq:Be_Axial_width}
\begin{split}
\Gamma^{\text{Be}}_{X,A} = 
&\frac{\left[w_0(\varepsilon^A_p+\varepsilon^A_n) + w_1(\varepsilon^A_p-\varepsilon^A_n)\right]^2 
\, {C^2_{{A},{\rm Be}}} \, \Lambda^2}
{(w_0^2 + w_1^2)\, 32\pi m_{N_\ast}^2} \\
&\times p_{X} \left(1+\frac{p_{X}^2}{3m_X^2} \right)
\end{split}
\end{equation}
and 
\begin{equation}
\label{eq:Be_vector_width}
\begin{split}
\Gamma^{\text{Be}}_{X,V} = 
&\left[w_0(\varepsilon^V_p+\varepsilon^V_n) + w_1(\varepsilon^A_p-\varepsilon^A_n)\right]^2 \\
&\times \frac{\alpha \, {C^2_{V,\text{Be}}}}{(w_0^2 + w_1^2)\, 12 \Lambda^2} \, p_{X}^3,
\end{split}
\end{equation}
where $w_0,w_1$ are coefficients that describe the excited nuclear state which we now discuss.

Due to the isospin mixing and breaking of the physical excited nuclear state of beryllium, care must be taken when matching the nucleus EFT to the nucleon EFT. The nucleon level EFT describing the transition is thus composed of a linear combination of currents from Eqs.~\eqref{eq:currentsv0}-\eqref{eq:currentsa1}. In particular, for vector coupled $X$ the interaction is written as $w_0(\epsilon_p^V + \epsilon_n^V)e J_\mu^0X^\mu + 
w_1(\epsilon_p^V - \epsilon_n^V)e J_\mu^1X^\mu$ where $w_0,w_1$ are the coefficients for isospin conserving and violating currents, respectively, and the interaction for an $X$ with axial-vector couplings is $w_0(\epsilon_p^A + \epsilon_n^A) J_{\mu,5}^0X^\mu + w_1(\epsilon_p^A - \epsilon_n^A) J_{\mu,5}^1X^\mu $. For the physical ${\rm Be}^8(18.15)$ state, the coefficients were found through a quantum Monte Carlo simulation to be ~\cite{Pastore:2014oda},
\begin{equation}
w_1 = -\alpha_1M1_{1,T=1} +\beta_1\kappa M1_{1,T=1}, w_0 = \beta_1 M1_{1,T=0},
\end{equation}
with $\alpha_1 = 0.21(3),\beta_1=0.98(1),\kappa=0.549,M1_{1,T=1}=0.767(9)\mu_N,M1_{1,T=0}=0.014(1)\mu_N$ describe the isospin mixing and breaking of the excited state, and $\mu_N$ is the Bohr magneton. 

We match the nucleus EFT amplitude to that of the the nucleon level to fit the Wilson coefficients. Because the photon also decays through a vector interaction and is thus proportional to $C_{V,{\rm Be}}$, it can be removed by relating the decay width to $X$ to the measured decay width of the photon following Sec.~4 of Ref~\cite{Feng:2016ysn} 
\begin{equation}
\label{eq:BeVectorDecay}
    \frac{\Gamma_{X,V}^{\rm Be}}{\Gamma\gamma} = \left(\frac{(e\epsilon_p^V - e\epsilon_n^V)w_1 + (e\epsilon_p^V + e\epsilon_n^V)w_0}{e(w_1+w_0)}\right)^2\frac{|\vec{k}_X|^3}{|\vec{k}_\gamma|^3},
\end{equation}
where $\vec{k}_{X,\gamma}$ are the decay product 3-momenta fixed by kinematics and $\Gamma_\gamma=1.9$ eV~\cite{nndc}.

However, the Wilson coefficient $C_{A,{\rm Be}}$ for an $X$ with axial-vector couplings cannot be obtained this way. In Ref.~\cite{Kozaczuk:2016nma} the nuclear matrix elements for axial-vector couplings were obtained using $N$-body simulations. The decay width was found to be 
\begin{equation}
\label{eq:BeAxialVectorDecay_isoscalar}
\begin{split}
\Gamma_{X,A}^{\rm Be} &= \frac{|\vec{k}_X|}{18\pi}\left(2+\frac{E_X^2}{m_X^2}\right) \\
&\quad \times 
\bigl(\epsilon_n^A\langle0|\sigma^n|{\rm Be}^8(17.64)\rangle
+ \epsilon^A_p\langle0|\sigma^p|{\rm Be}^8(17.64)\rangle\bigr)^2,
\end{split}
\end{equation}
with transition matrix elements $\langle0|\sigma^n|{\rm Be}^8(17.64)\rangle=-0.073(29)$ and $\langle0|\sigma^p|{\rm Be}^8(17.64)\rangle = 0.102(28)$

We repeat this analysis for the exotic decay of the the mostly iso-scalar state, ${\rm Be}(18.15)$. Because it has the same $J^P$ quantum numbers as the mostly iso-vector state above, the calculation is very similar, but with some substitutions for the coefficients and matrix elements. For the nucleon EFT, the $w_1,w_0$ coefficients become~\cite{Feng:2016ysn}
\begin{equation}
    w_1=\beta_1M1_{1,T=1}+\alpha_1\kappa M1_{1,T=1},~w_0 = \alpha_1 M1_{1,T=0}.
\end{equation}

For an $X$ with vector couplings, the partial width can be obtained from~\cref{eq:Be_vector_width} and $\Gamma_\gamma = 15 $ eV. For an $X$ with axial-vector couplings, the partial width is \cref{eq:Be_Axial_width} and the derivation for the matrix element is similar to \cref{eq:BeAxialVectorDecay_isoscalar},
\begin{equation}
\label{eq:BeAxialVectorDecay_isovector}
\begin{split}
\Gamma_X &= \frac{|\vec{k}_X|}{18\pi}\left(2+\frac{E_X^2}{m_X^2}\right) \\
&\quad \times 
\bigl(\epsilon_n^A\langle0|\sigma^n|{\rm Be}^8(18.15)\rangle
+ \epsilon^A_p\langle0|\sigma^p|{\rm Be}^8(18.15)\rangle\bigr)^2,
\end{split}
\end{equation}
with the transition matrix elements $\langle0|\sigma^n|{\rm Be}^8(18.15)\rangle=-0.132(33)$ and $\langle0|\sigma^p|{\rm Be}^8(18.15)\rangle = -0.047(0.29)$~\cite{Kozaczuk:2016nma}. 

For each physical beryllium state we calculate the total width using \cref{eq:Be_Axial_width,eq:Be_vector_width,eq:BeAxialVectorDecay_isoscalar,eq:BeAxialVectorDecay_isovector,eq:BeVectorDecay}. When fitting, we propagate and include theoretical uncertainties on the axial-vector matrix element, which are then added in quadrature to the experimental uncertainties.

\subsection{Helium} 

As discussed above, there are two relevant excited states for helium, $^4{\rm He}(21.01)$ and $^4{\rm He}(20.21)$, which have $J^P$ quantum numbers of $0^-$ and $0^+$, respectively. The experiment was run at an intermediate energy, such that both excited states would be produced. The relevant quantity is then the total $X$ production cross section $\sigma_X$, related to the branching fraction of each helium state to $X$ by $\sigma_X=\sigma_-{\rm BF}(0^-\rightarrow X) + \sigma_+{\rm BF}(0^+\rightarrow X)$, where $\sigma_\pm$ is the production cross section for the $0^\pm$ state. The ATOMKI collaboration reports the ratio of the cross section for $X$ production to that of the E0 transition, which can be defined in a similar way, $\sigma_{\rm E0}=\sigma^+ (\Gamma_{\rm E0}/\Gamma_+)$. The production cross section to create the $0^\pm$ state from a beam corresponding to a center of mass energy $E_{\rm CM}$, normalized to the E0 transition, $\sigma_X / \sigma_{\rm E0}$ can be written
\begin{equation}
\label{eq:sigmax_sigmaE0}
\begin{aligned}[t]
\frac{\sigma_X}{\sigma_{\rm E0}} 
&= \frac{\Gamma(0^+\!\rightarrow X)}{\Gamma_{\rm E0}} \\[6pt]
&\quad + \frac{0.76 \, \Gamma_-}{\Gamma_+}
    \left(
    \frac{(E_{\rm CM}^2 - M_+^2)^2 + M_+^2 \Gamma_+^2}
         {(E_{\rm CM}^2 - M_+^2)^2 + M_-^2 \Gamma_-^2}
    \right)
    \frac{\Gamma(0^-\!\rightarrow X)}{\Gamma_{\rm E0}},
\end{aligned}
\end{equation}
with $\Gamma_\pm$ the total width for the $0^\pm$ state with mass $M_\pm$.

Now, the partial width $\Gamma(0^\pm\rightarrow X)$ can be calculated. If $X$ has only vector couplings, it must be produced from the $0^+$ state, while axial-vector coupled $X$ are necessarily produced from $0^-$. Both decays are mediated by an EFT operator of the form $N_0^\dagger X^\mu\partial_\mu N_*$. The resulting decay width for the isospin conserving decay is thus similar in each case, with differences arising from the different final state momenta, and a factor $\alpha$ is included for the vector case per convention, so that
\begin{align}
\Gamma(0^+\rightarrow X )&=(\epsilon_p^V+\epsilon_n^V)^2\frac{\alpha C^2_{V,{\rm He}}}{8m_X^2}p_X^3, \\
\Gamma(0^-\rightarrow X )&=(\epsilon_p^A+\epsilon_n^A)^2\frac{ C^2_{A,{\rm He}}}{32\pi m_X^2}p_X^3.
\end{align}
The Wilson coefficient is matched for each case by using the observed E0 decay rate for the vector state which has the same dependence on the Wilson coefficient. For the axial-vector, we use the resulting nuclear matrix elements reported from Ref.~\cite{Horiuchi:2009zza} and used for analysis of the $X$ in Ref.~\cite{Barducci:2022lqd}. For theory uncertainties, two benchmark values of 50\% and 100\% are assumed on the nuclear matrix element.

\subsection{Carbon} 

The excited state of carbon, $^{12}{\rm C}(17.23)$, decays via an isovector transition. The leading operator for an $X$ with (axial-)vector transition is of dimension (5) 3. The effective operator at the nucleus level for the isospin changing decay is then
\begin{equation}
\label{eq:CarbonOperator}
\begin{aligned}[t]
& C_{A,C}\!\left(\frac{\epsilon_p^A-\epsilon_n^A}{2}\right)\!\Lambda 
  N_0^\dagger N_*^\mu X_\mu \\[6pt]
&\quad + \frac{C_{V,C}}{\Lambda} \!\left(\frac{e\epsilon_p^V-e\epsilon_n^V}{2}\right)\! N_0^\dagger 
  \epsilon^{\mu\nu\alpha\beta} (\partial_\mu N_{*^\nu})(\partial_{\alpha}X_\beta),
\end{aligned}
\end{equation}
where the first (second) term mediates transitions for $X$ with (axial-)vector couplings. As in the beryllium case, the spin averaged squared matrix elements do not interfere, and the total decay width $\Gamma(C\rightarrow X)$ is thus the sum of the axial-vector and vector contributions
\begin{multline}
\Gamma(C \rightarrow X) = \Gamma_{X,V}^{\rm C} + \Gamma_{X,A}^{\rm C} =\\[4pt]
(\epsilon_p^A - \epsilon_n^A)^2 
    \frac{C_{A,C}^2}{48\pi\Lambda^2} p_X^{3} \\[8pt]
+ (\epsilon_p^V - \epsilon_n^V)^2
    \frac{\alpha C_{V,C}^2 \Lambda^2}{8 m_{N_*}^2} 
    p_X \left(1 + \frac{p_X^2}{3 m_X^2}\right).
\end{multline}

Each Wilson coefficient is again matched to the decay rate of the photon for the vector case, and to an analytical calculation of nuclear structure for the axial-vector case. For the vector case, there is some subtlety as the operator in \cref{eq:CarbonOperator} is not gauge-invariant if $X$ is replaced by a photon---however, the gauge invariant photon operator can be related to the vector operator in \cref{eq:CarbonOperator}. We refer the reader to Sec.~7B of Ref.~\cite{Feng:2020mbt}, and conclude that the vector contribution to the exotic carbon decay can be related to the photon decay width as
\begin{equation}
    \frac{\Gamma_{X,V}^{\rm C}}{\Gamma_{\gamma}}=\frac{3}{2}(\epsilon_p^V-\epsilon_n^V)^2\frac{p_X}{p_\gamma}E_X^2\left(1 + \frac{p_X^2}{3m_X^2}\right).
\end{equation}

Turning to the matching of the Wilson coefficient, $C_{A,C}$, there is not much work on the evaluation of the appropriate matrix elements. In Ref~\cite{Mommers:2024qzy}, the authors evaluated the partial decay width for the excited carbon nucleus to decay to either an axial-vector $X$ or a photon using a nuclear shell model. The authors found that the nuclear effects cancel in the ratio
\begin{equation}
    \frac{\Gamma_{X,A}^{\rm C}}{\Gamma_\gamma} = \frac{1}{4}\left[1 - \left(\frac{m_X}{17.23~{\rm MeV}}\right)^2\right]^{3/2}\frac{1}{4\pi\alpha}(\epsilon_p^A - \epsilon_n^A)^2,
\end{equation}
which is used for evaluating the Wilson coefficient in exchange for the measured partial decay width of the $C\rightarrow\gamma$ decay. However, using the result from Ref~\cite{Mommers:2024qzy} comes with the important caveat that they calculate a photon decay width of 251 eV that is significantly larger than the experimentally measured value of 44 eV, raising a question on the reliability of the method. Therefore, as arbitrary theoretical uncertainty benchmarks, we implement a fractional uncertainty of 50\%, or 100\% on the resulting width $\Gamma_{X,A}^{\rm C}$.

\bibliography{references}
\end{document}